\documentclass[10pt,twocolumn,letterpaper]{article}

\usepackage[pagenumbers]{cvpr} 

\definecolor{cvprblue}{rgb}{0.21,0.49,0.74}
\usepackage[pagebackref,breaklinks,colorlinks,allcolors=cvprblue]{hyperref}
\usepackage{xcolor}
\usepackage{amsmath}
\usepackage[ruled,linesnumbered]{algorithm2e}
\usepackage{multirow}
\usepackage{hyperref}
\usepackage{textcomp}
\usepackage{makecell}

\def\confName{CVPR}
\def\confYear{2025}

\title{
ArcPro: Architectural Programs for Structured 3D Abstraction of Sparse Points
}

\author{
Qirui~Huang{\textsuperscript{1,2}},
Runze~Zhang{\textsuperscript{1}},
Kangjun~Liu{\textsuperscript{2}},
Minglun~Gong{\textsuperscript{3}},
Hao~Zhang{\textsuperscript{4}},
Hui~Huang{\textsuperscript{1}\thanks{Corresponding author}}\\
\textsuperscript{1}CSSE, Shenzhen University 
\textsuperscript{2}Pengcheng Laboratory 
\textsuperscript{3}University of Guelph
\textsuperscript{4}Simon Fraser University
}

\definecolor{freshgreen}{RGB}{34, 200, 34}  
\definecolor{skyblue}{RGB}{70, 130, 180}    
\definecolor{peach}{RGB}{255, 160, 122}     
\definecolor{peachc}{RGB}{255, 150, 150}     
\definecolor{lavender}{RGB}{181, 126, 220}  
\definecolor{mint}{RGB}{152, 251, 152}      
\definecolor{coral}{RGB}{240, 128, 128}     
\definecolor{coolgray}{RGB}{119, 136, 153}  
\definecolor{myPurple}{RGB}{138,43,226}

\begin{document}

\twocolumn[{%
\renewcommand\twocolumn[1][]{#1}%
\maketitle
    \centering
    \includegraphics[width=1\textwidth]{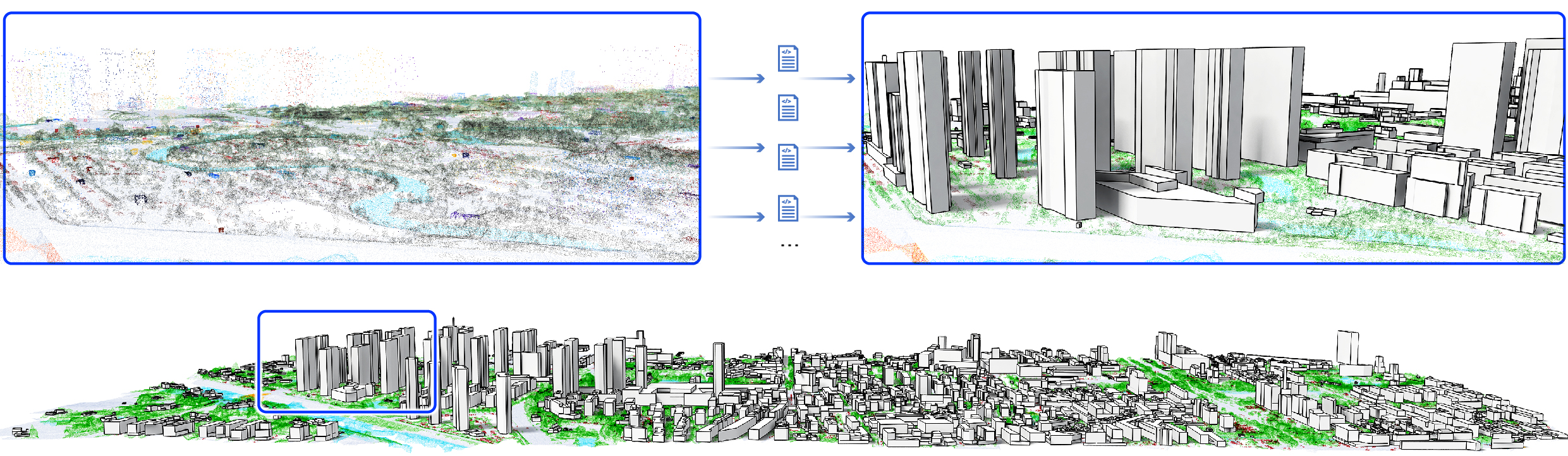}
    \vspace{-2em}
    \captionof{figure}{Structured 3D abstraction (bottom and top right for a zoom-in) in the form of {\em architectural programs\/}, obtained by our method ArcPro, which takes as input an extremely sparse point cloud (only $\approx$300 points) segment over each building. Despite such low-density, non-uniform, and noisy inputs, our method produces clean, low-face-count meshes that structurally conform to the real building objects. ArcPro can process 1,090 buildings over an area of 2.92 $km^2$ in approximately 37 seconds on a single 4090 GPU.}
    \vspace{2mm}
\label{fig:teaser}
}]

\maketitle

\renewcommand{\thefootnote}{}
\footnotetext[1]{* Corresponding author.}

\begin{abstract}
\vspace{-5pt}

We introduce ArcPro, a novel learning framework built on architectural programs to recover structured 3D abstractions from highly sparse and low-quality point clouds.
Specifically, we design a domain-specific language (DSL) to hierarchically represent building structures as a program, which can be efficiently converted into a mesh.
We bridge feedforward and inverse procedural modeling by using a feedforward process for training data synthesis, allowing the network to make reverse predictions.
We train an encoder-decoder on the points-program pairs to establish a mapping from unstructured point clouds to architectural programs, where a 3D convolutional encoder extracts point cloud features and a transformer decoder autoregressively predicts the programs in a tokenized form.
Inference by our method is highly efficient and produces plausible and faithful 3D abstractions.
Comprehensive experiments demonstrate that ArcPro outperforms both traditional architectural proxy reconstruction and learning-based abstraction methods.
We further explore its potential to work with multi-view image and natural language inputs.
Project page: \url{https://vcc.tech/research/2025/ArcPro}.

\end{abstract}
    
\vspace{-10pt}

\section{Introduction}
\label{sec:intro}

Efficient extraction of {\em structured\/} 3D representations from unstructured architectural scene acquisitions such as point clouds is crucial for urban modeling, planning, and spatial computations in applications like autonomous navigation, augmented reality, and digital twins~\cite{biljecki2015applications}. However, building a mapping from unstructured point clouds to meshes representing potential architectural entities poses two main challenges.
First, raw point clouds from aerial or ground scanners often contain missing data and noise, while point clouds obtained from acquired images, e.g., through structure-from-motion (SfM), may even be of lower quality with additional sparsity and non-uniformity.
Such low data qualities necessitate the integration of prior knowledge to identify architectural features, since traditional methods relying on constraints such as the manhattan~\cite{li2016boxfitting} or planar hypothesis~\cite{bauchet2020kinetic} are inadequate for complex structures. Second, direct mapping from sparse points to the mesh space is challenging due to the coupling of geometric data and connectivity relationships, where an overly flexible representation can easily lead a neural model to exhibit excessive sensitivity to noise and other data artifacts from the input.

In this paper, we introduce a {\em program-based\/} learning framework to recover structured 3D abstractions from low-quality, unstructured building point clouds. Our core idea is to design a domain-specific language (DSL), called {\em architectural programs\/} or ArcPro for short, which serves as an intermediate representation, dividing the 3D abstraction problem into two steps: mapping point clouds to programs and mapping programs to meshes. Our program representation for architectural models has its roots in classical graphics methods for procedural and grammar-based city and building modeling~\cite{parish2001procedural,muller2006procedural} and offers three advantages:

\begin{itemize}
    \item  Our DSL models building {\em hierarchically\/} using architectural trees, which is a compact and intuitive representation conforming to architectural design principles. 
    \item With a controlled representational capacity, our DSL
can cover most prevalent architectural structures without overfitting to noisy or incomplete data.
    \item The procedural nature of our programs allows easy data generation, which, when coupled with data augmentation via point sampling, allows us to create large volumes of program-point cloud pairs to train our mapping network.
\end{itemize}

We train an encoder-decoder on the points-program pairs to establish a mapping from unstructured point clouds to architectural programs, where a 3D convolutional encoder extracts point cloud features and a transformer decoder autoregressively predicts the programs in a tokenized form to minimize a next-token prediction loss. To establish a bijective mapping between our architectural program and a token sequence for the transformer, an architectural tree is serialized into a sequence of nodes through breadth-first traversal, which is further converted into a geometrically equivalent program for mesh conversion. To ensure syntactic correctness of the predicted tokens, we design a masking strategy with the aid of a finite state machine (FSM) to prevent erroneous tokens that would introduce syntax errors based on the context of the preceding token sequence.

\begin{figure*}[!ht]
    \centering    \includegraphics[width=\textwidth]{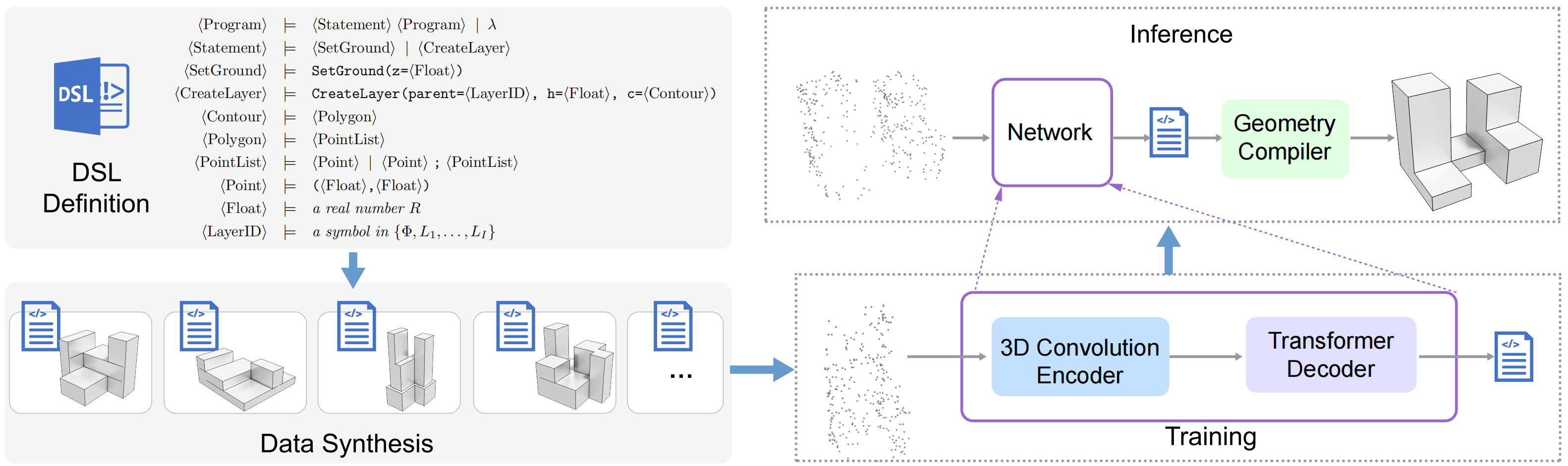}
    \caption{
Our architectural programs, a DSL defined using Backus-Naur Form, and the ArcPro method for structured 3D abstraction from point clouds.
Procedural generation synthesizes paired programs and 3D meshes, from which point clouds are sampled to create input-output pairs.
The network, consisting of a 3D convolutional encoder and a transformer decoder, is trained to autoregressively predict a program in tokenized format, which is then compiled into a 3D mesh as a structured abstraction of the input during inference.
}
    \label{fig:method}
\end{figure*}

During inference, the trained network uses the input point cloud as a condition to generate an architectural program.
Then, we employ a learning-free interpreter, akin to a geometry compiler, to translate the predicted program into a mesh -- a structured 3D abstraction of the input; see Fig.~\ref{fig:method} for our method pipeline.

Our work builds on the recent successes on learning visual programs and neuralsymbolic representations for CAD shapes and other visual concepts~\cite{ritchie2023neurosymbolic,gupta2023visual}.
Our main contributions can be summarized as follows.
\begin{itemize}
    \item To the best of our knowledge, ArcPro is the first program-based method for structured representation learning from sparse architectural point clouds. Prior inverse procedural models in this domain are based on either optimization~\cite{mathias2011procedural} or template instantiation~\cite{toshev2010detecting}.
    \item We connect feedforward and inverse procedural modeling by applying a feedforward process to synthesize training data, enabling the network to make reverse predictions.
    \item Inference by our ArcPro method is highly efficient and produces plausible structured 3D architectural abstractions conforming to the reference despite low-quality point cloud inputs with sparsity, noise, non-uniformity, and incompleteness, as shown in Fig.~\ref{fig:teaser}.
\end{itemize}

Comprehensive experiments demonstrate that ArcPro outperforms existing architecture proxy reconstruction and learning-based 3D abstraction methods. We also analyze the performance of our method in various low-quality point cloud scenarios and show its potential with other modalities, e.g., multi-view images and natural languages.

\section{Related Works}
\label{sec:rw}

\paragraph{Architectural proxy reconstruction.}
Architectural proxy reconstruction aims to automatically rebuild the main structures of buildings from unstructured point clouds. Existing methods typically follow a pipeline of primitive detection and assembly. For instance, \citet{chauve} propose an adaptive 3D space decomposition using planar primitives, generating a watertight mesh through Delaunay triangulation. \citet{LinH} fit parametric building blocks to LiDAR data for building reconstruction. Polyfit~\cite{nan2017polyfit} apply optimization techniques based on integer programming to approximate building geometries. KSR~\cite{bauchet2020kinetic} develop a more efficient algorithm to combine detected primitives. However, these methods require dense, high-quality point clouds to ensure that primitive extraction algorithms~\cite{ransac,RApter} can yield reasonable primitives for surface assembly. As a result, they often fail to produce plausible results when working with incomplete or noisy data. The recent ProxyRecon~\cite{ProxyRecon24} avoids using primitive detection for proxy reconstruction but struggles with common non-convex building structures.

\vspace{-10pt}

\paragraph{Learning 3D structures and abstractions.}
Shape abstraction aims to capture the underlying structure which approximates complex shapes.
Most approaches focus on predicting the parameters of predefined geometric primitives.
These primitives can explicitly be planes~\cite{chen2020bsp}, cuboids~\cite{yang2021unsupervised}, and superquadrics~\cite{paschalidou2019superquadrics}.
The primitives may also be represented implicitly~\cite{niu2022rim, shuai2023dpf}, which requires conversion to meshes using techniques like Marching Cubes~\cite{barrow1977parametric}.
However, this conversion often results in meshes that are not very clean.
Neurosymbolic 3D shape modeling~\cite{ritchie2023neurosymbolic} offers a way to represent clean geometry through programs.
Researchers have developed various domain-specific languages (DSLs) to formulate these programs~\cite{jones2020shapeassembly, jones2023shapecoder, avetisyan2024scenescript} and have also used CAD construction commands~\cite{wu2021deepcad, zhou2023cadparser, xu2022skexgen}, both tailored to different datasets or scenarios~\cite{mo2019partnet, koch2019abc, willis2021fusion}.
These approaches reduce complexity by narrowing the solution space to a more compact program space, which inspires us to infer architectural programs from sparse points.

\vspace{-10pt}

\paragraph{Procedural bulding models.}
Feedforward programs aim to develop procedural grammars that enable users to write rules to produce architectures, which are both interpretable and editable~\cite{parish2001procedural,muller2006procedural}. However, crafting rule documents from scratch is often challenging for most users, who typically need to adjust existing rule templates.
Consequently, there has been an interest in Inverse Procedure, which focuses on reconstructing procedural representations from input data such as point clouds~\cite{mathias2011procedural, toshev2010detecting}.
These methods rely on carefully designed algorithms to embed architectural priors and an initial RANSAC plane extraction, which is prone to failure under sparse conditions.
Our method construct a simple feedforward procedural process, and learn the deep network to establish the inverse mapping.

\section{Overview}
\label{sec:overview}

Our goal is to recover the primary 3D structure of a building from a sparse architectural point cloud, even in challenging cases where the data is noisy, non-uniform, and incomplete. The key idea of ArcPro is to represent the architectural structure as a \textit{Program}, which serves as an intermediary between the input point cloud and the output mesh; see Figure~\ref{fig:method}. To achieve this, we tackle two key challenges:

\begin{enumerate}
    \item \textbf{Domain Specific Language (DSL)}: How to design a DSL that effectively represents architectural structures?
    \item \textbf{Training Data}: How to acquire suitable training data? 
\end{enumerate}

The key challenge lies in bridging the gap between the hierarchical nature of architectures and the linear representation used in programs. We model architectural structures using trees. An architectural tree is serialized into a sequence of nodes through breadth-first traversal, which is further converted into a geometrically equivalent program. Finally, the program is processed into meshes, similar to how compilers translate code into executable formats.

To generate training data, we synthesize architectural trees using procedural generation and convert them into programs, allowing us to produce large-scale datasets.  By sampling point clouds from the converted meshes, we obtain paired datasets of input point clouds and their corresponding ground truth programs. We then take the point clouds as conditional input to our model, employing an autoregressive approach with next-token prediction loss to output the program in a tokenized format.

\section{Method}
\label{sec:method}

In order to recover the primary 3D structure of a building from a sparse architectural point cloud, we aim to model the conditional probability distribution $p(\mathbf{Y}\mid\mathbf{X})$,  
where $\mathbf{X} = \{\mathbf{x}_1, \dots, \mathbf{x}_N\}$ is the input point cloud consisting of $N$ points. Each point $\mathbf{x}_i \in \mathbb{R}^3$ denotes the 3D coordinates of the $i$-th point.  
The output $\mathbf{Y}$ is the underlying primitive geometry represented as a mesh.

To achieve this, we introduce the program $\mathbf{P}$ as an intermediate representation.
We use a learnable network $\theta$ to model ${p_\theta}(\mathbf{P} \mid \mathbf{X})$.  
The network is trained using supervised learning with paired data $(x, p)$, where $x \in \mathbf{X}$ represents an input point cloud, and $p \in \mathbf{P}$ corresponds to the ground truth program.
The conditional distribution $p(\mathbf{Y} \mid \mathbf{P})$ is modeled as a deterministic function $\mathcal{G}$ that maps the program $\mathbf{P}$ to the corresponding mesh: $\mathbf{Y} = \mathcal{G}(\mathbf{P})$.
Thus, we transform the solution space from the mesh space to a more compact program space:
\begin{equation}
    p_\theta(\mathbf{Y} \mid \mathbf{X}) = \int \mathbb{I}\left[\mathbf{Y} = \mathcal{G}(\mathbf{P})\right] \, p_\theta(\mathbf{P} \mid \mathbf{X}) \, d\mathbf{P},
    \label{eq:decomposition_2}
\end{equation}
where \( \mathbb{I}[\cdot] \) is the indicator function, which equals 1 if the condition inside holds (\ie, \( \mathbf{Y} = \mathcal{G}(\mathbf{P}) \)) and 0 otherwise.

\begin{figure}[tbp]
    \centering    \includegraphics[width=0.48\textwidth]{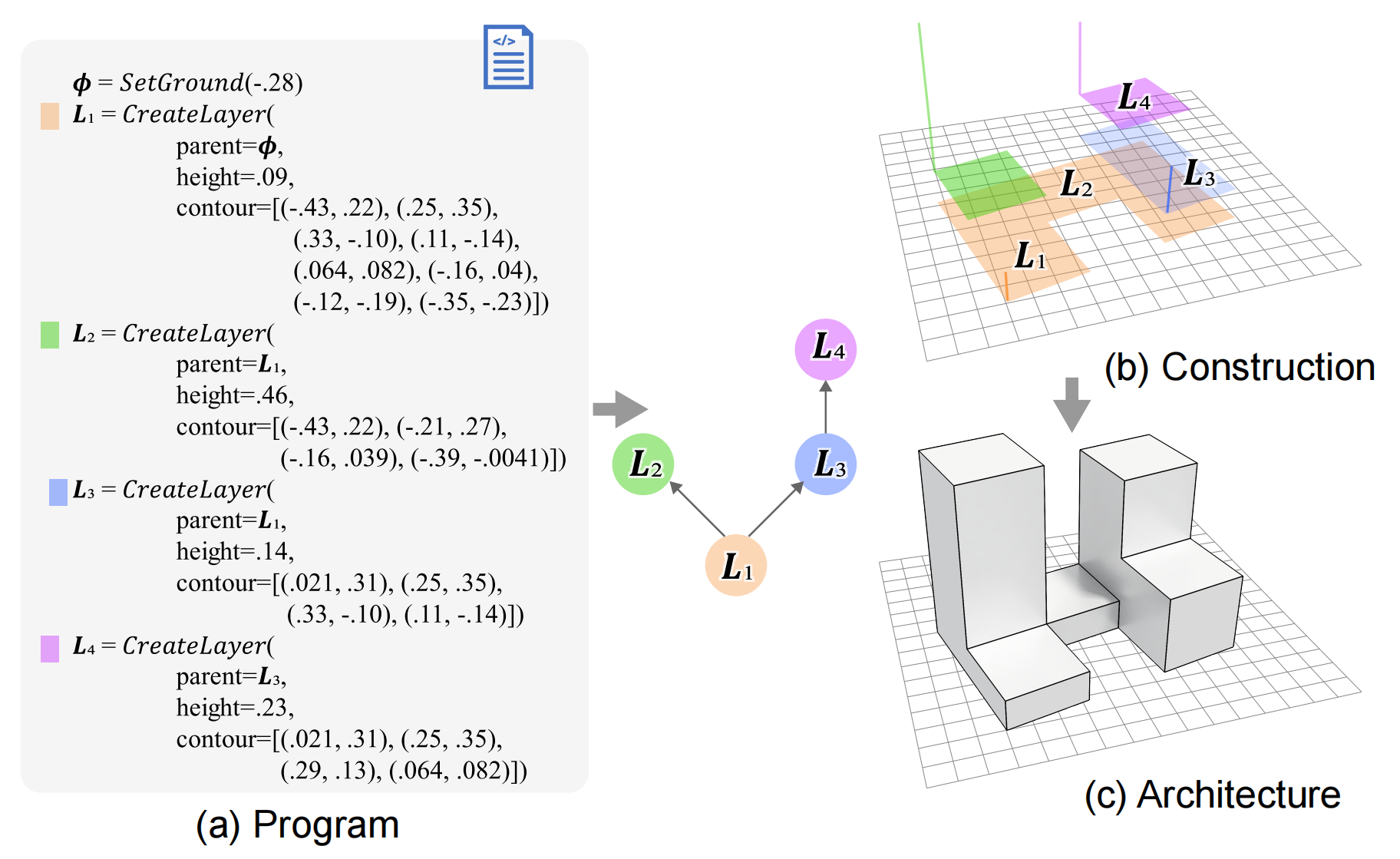}
    \caption{
The geometry compilation process transforms the program into architectural meshes by constructing an architectural tree that encodes layer heights, contours, and spatial relationships.}
    \label{fig:vis_program}
\end{figure}

\subsection{Domain-Specific Language}
\label{sec:dsl}
We assume the architectural structure is a geometric body above the ground plane. The DSL aims to encode two key pieces of information: the ground height and the shape of the building profile.
Figure~\ref{fig:vis_program} illustrates the construction process from program to architecture.

\vspace{-10pt}

\paragraph{Statement for localization.}
In real-world scenarios, architectural point clouds are in a global coordinate system with one axis perpendicular to the ground, assumed to be the $z$-axis.
Although the exact ground location is unknown, it can be specified by a single z-coordinate.
To enable the network to predict it, we design the following statement:
$$\Phi = \texttt{SetGround}(z=z_{\Phi}),$$
where $\Phi$ is the ground plane at $z$-coordinate $z_{\Phi}$.

\vspace{-10pt}

\paragraph{Architectural tree.}
We model the architectural structure as a rooted tree \( T = (V, E) \), where \( V \) represents the set of nodes, and \( E \subset V \times V \) defines the parent-child relationships among nodes. Geometrically, we express $V$ as \( V = \bigcup_{i=1}^{I}L_i \), where each $L_i$ represents a layer, and \( E \) captures the spatial hierarchy among them. Each layer \( L_i \) is a geometric column characterized by two attributes: $L_i = (h_i, c_i)$, where \( h_i \in \mathbb{R}^+ \) denotes the height, and \( c_i \subset \mathbb{R}^2 \) specifies the 2D contour in the form of a polygon.

\vspace{-10pt}

\paragraph{Statement conversion.}
We serialize the architecture tree \( T = (V, E) \) into a node sequence using breadth-first search:
$$\langle L_{\sigma(1)}, L_{\sigma(2)}, L_{\sigma(3)}, \dots, L_{\sigma(I)} \rangle$$
where \( \sigma: \{1, 2, \dots, I\} \to \{1, 2, \dots, I\} \) is a permutation function that defines the order in which the nodes are traversed.
To convert this node sequence into statements, we design a statement in our DSL: \texttt{CreateLayer}. The syntax is defined as follows:
$$L_i = \texttt{CreateLayer}(parent=L_j, h=h_i, c=c_i),$$
where \( L_j \) is the parent of \( L_i \), and \( h_i \) and \( c_i \) are its attributes. If \( L_j = \Phi \), it means that \( L_i \) is the root node.
Each node in the sequence corresponds to a statement in the program, thereby translating the hierarchical structure of the architecture into a linear programmatic representation.

\begin{figure}[tbp]
    \centering    \includegraphics[width=0.48\textwidth]{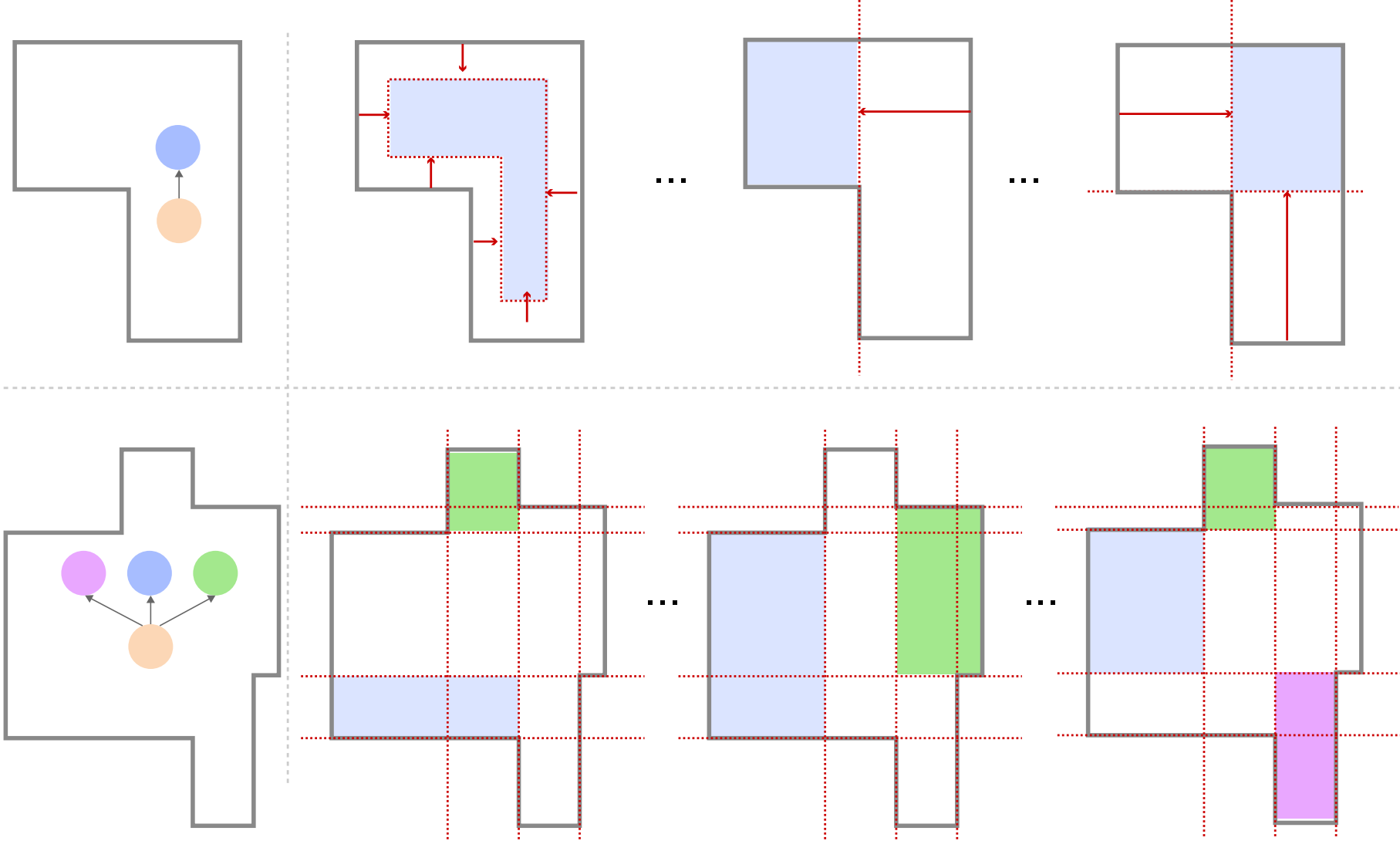}
    \caption{Child nodes contour generation: single child (top) and multiple children (bottom). Left shows parent node contour; right shows possible child node contours.}
    \label{fig:vis_pcg}
\end{figure}

\subsection{Procedural Generation}
\label{pcg}

The next challenge is to obtain a large-scale training dataset of programs based on the format defined above. Since we have established a connection between architecture trees and programs, this challenge translates into large-scale synthesis of architecture trees.

We model the synthesis process as procedural generation.
Specifically, we iteratively generate a tree where, in each iteration, a leaf node is randomly selected to spawn child nodes. The height of new child nodes are randomized within a specified range, and 2D contours are sampled according to the following two scenarios:

\begin{itemize}
    \item \textbf{Initialize root node} \(p(c_{L_1})\). To enhance realism, the root node's contour is sampled from the BingMaps~\cite{microsoft_global_ml_building_footprints} dataset, which contains a large collection of real-world building footprints. This provides a diverse set of contours for the ground levels of architectural models.

    \item \textbf{Add child nodes} \(p(\{c_{L'_m}\} \mid c_L)\). This step involves determining \( M \) subregions within the parent node’s contour \( c_L \): 1) \( M = 1 \): Generate a single child contour \( c_{L'} \) by contracting selected edges of \( c_L \) inward by random distances. 2) \( M > 1 \): Extend edges of \( c_L \) to perform planar bisection, splitting the interior into cells. Sample \( M \) blocks as unions of adjacent cells to form subregions \(\{c_{L'_m}\}\) with area and distance constraints. The specific examples are shown in Figure~\ref{fig:vis_pcg}.

\end{itemize}

\begin{table}[t]
  \centering
  \resizebox{0.48\textwidth}{!}{
    \begin{tabular}{c}
      \toprule
      Statement and Its Tokens \\
      \midrule
      $\Phi = \texttt{SetGround}(z=z_{\Phi})$ \\
      $\langle\Phi\rangle\langle\text{SetGround}\rangle[z_{\Phi}]\langle\text{/z}\rangle$ \\
      \midrule
      $L_i = \texttt{CreateLayer}(parent=L_j, h=h_i, o=o_i)$ \\
      $\langle L_i\rangle\langle\text{CreateLayer}\rangle\langle L_j\rangle[h_i]\langle\text{/h}\rangle[x_i^1][y_i^1]\langle\text{/p}\rangle\dots[x_i^n][y_i^n]\langle\text{/p}\rangle$ \\
      \bottomrule
    \end{tabular}
  }
  \caption{The tokenization rules for each statement.}
  \label{tab:tokenization}
\end{table}

\subsection{Training and Inference}

\paragraph{Tokenization.} The goal of tokenization is to establish a bijective mapping between a program and a sequence of integer tokens, which serves as the data format that the network can process. We classify token types into two categories: \textit{Numeric Tokens} and \textit{Non-Numeric Tokens}. Numeric Tokens represent all numerical values, such as coordinates or height values, whereas Non-Numeric Tokens denote the syntax structure or label the nodes within the program. The tokenization rules for each statement are shown in Table~\ref{tab:tokenization}.
Notably, this tokenization scheme is distinct from traditional NLP methods like Byte-Pair Encoding (BPE).

\vspace{-10pt}

\paragraph{Network and loss function.}
Our network employs a simple yet effective encoder-decoder scheme. The input is a point cloud, and the output is a sequence of tokens that can be de-tokenized into a program. The training approach is autoregressive, relying solely on next-token prediction loss. Specifically, our encoder is a 3D sparse convolutional network that extracts features from the point cloud. These features are then flattened into a feature sequence for the decoder.
The Transformer decoder autoregressively predicts the next token, with point cloud features injected as conditional information through cross-attention.

\vspace{-10pt}

\paragraph{Syntax-constrained token sampling.} Our model predicts a sequence of tokens, which must be detokenized into a syntactically valid program. However, syntactic errors may arise during inference.
To this end, we propose a masking strategy that ensures the syntactic correctness of each predicted token during inference. This strategy constructs a finite state machine (FSM) model to guide token selection, masking tokens that would introduce syntax errors based on the context of the preceding token sequence.
For example, if the most recent token is $\langle\text{SetGround}\rangle$, the subsequent token must be a numeric value to represent $z_\Phi$. To enforce this constraint, the strategy masks all non-numeric tokens, allowing only valid candidates for the next token.
This approach integrates seamlessly with other sampling techniques, such as top-K, top-P, and beam search, without introducing conflicts or diminishing their effectiveness.

\vspace{-10pt}

\paragraph{Geometry refinement.}
During training, we apply a cross-entropy next-token prediction loss, leading to probabilistic token sampling during inference. Combined with a discrete coordinate representation, this can introduce imprecision, causing the model to miss identical coordinates for two points or misplace a point slightly off an edge to impact visualization quality.
To address this, we implement geometry refinement in post-processing. This method recursively adjusts each layer node’s contour by snapping points to the nearest points or edges on the parent node’s contour within a specified  threshold.
This approach not only enhances visualization clarity but also encodes spatial relationships between parent and child contours, enabling easy edits.

\section{Experiments}
\label{sec:exp}

\begin{figure*}[tbp]
    \centering    \includegraphics[width=\textwidth]{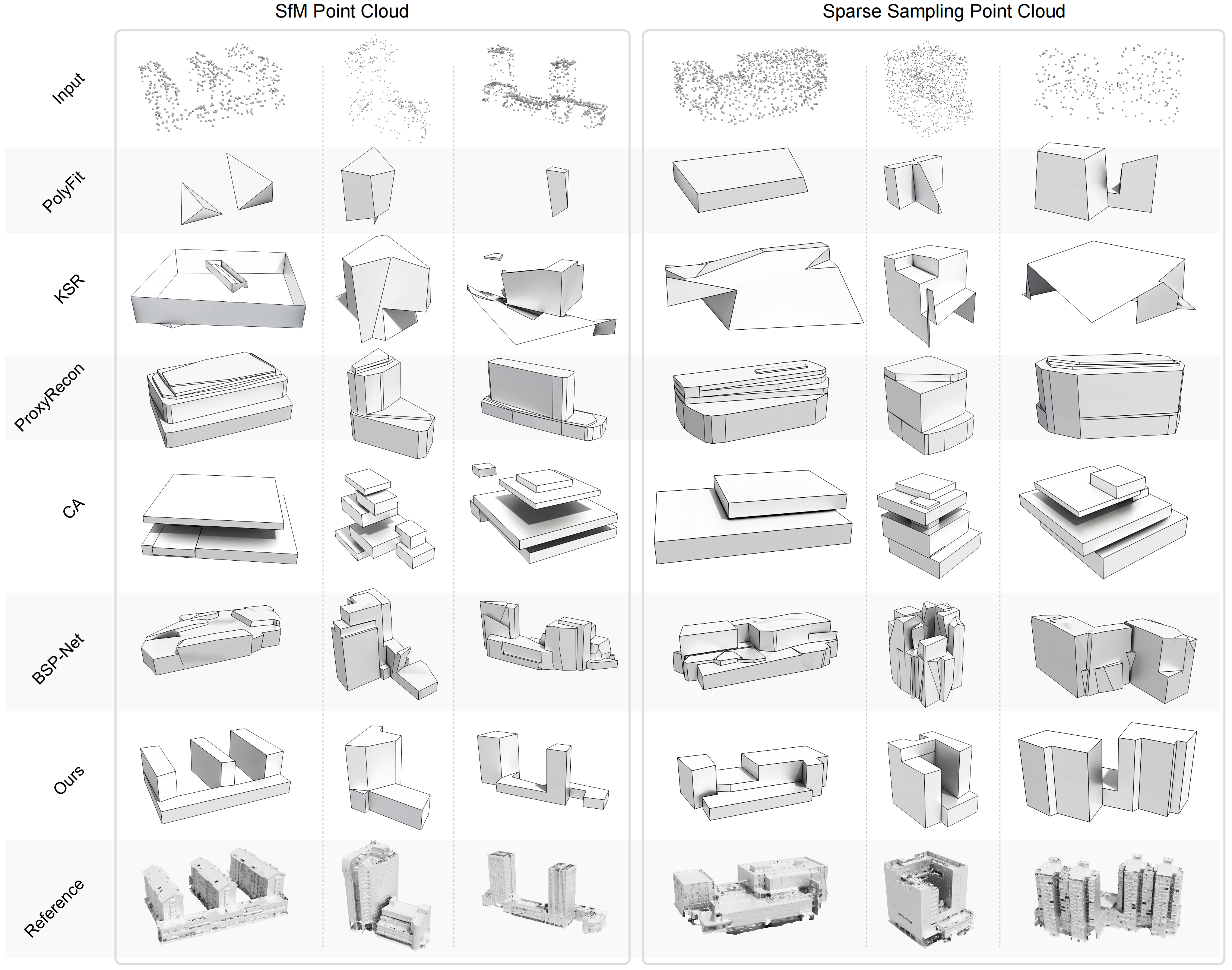}
    \caption{
    Qualitative comparison of our method with state-of-the-art (SOTA) methods on two evaluation datasets: the SfM and sparse sampling point clouds. The SOTA methods include traditional proxy reconstruction (PolyFit~\cite{nan2017polyfit}, KSR~\cite{bauchet2020kinetic}, ProxyRecon~\cite{ProxyRecon24}) and learning-based 3D abstraction (CA~\cite{yang2021unsupervised}, BSP-Net~\cite{chen2020bsp}). Our method outperforms all these alternatives. ArcPro's program representation balances the geometric primitives of cubes (as CA) and planes (as BSP-Net), enabling more efficient 3D abstraction of building structures.
    }
    \label{fig:comparison}
\end{figure*}

\subsection{Implementation}

\paragraph{Synthetic training dataset.}

Based on the procedural generation introduced in Section~\ref{pcg}, we use an architectural program synthesizer to generate training data online.
To bias the training data toward clean, well-formed geometries, we developed a validator to filter contours based on quality. Valid contours must be free of self-intersections, have interior angles in [20$^\circ$, 160$^\circ$], a longest-to-shortest edge ratio of less than 10, and an area between 15\% and 85\% of the parent contour's area.
For \( p(c_{L_1}) \), we cleaned 872,487 footprints from the Bing Maps~\cite{microsoft_global_ml_building_footprints} dataset to serve as contours for the first layer.  
For \( p(\{c_{L'_m}\} \mid c_L) \), our synthesizer randomly generates potential sub-contours based on \( c_L \) until all passing  the validator.  
We use \( \mathcal{G} \) to convert these synthetic programs into meshes and sample sparse points from them to form paired training data of programs and points.

\begin{table*}[]
\centering
\resizebox{\textwidth}{!}{
\begin{tabular}{ccccccclccccccc}
\toprule
\multirow{2}{*}{Method} & \multicolumn{6}{c}{SfM Point Cloud} &  & \multicolumn{6}{c}{Sparse Samping Point Cloud} & \multirow{2}{*}{User Study} \\ \cline{2-7} \cline{9-14}
                        & \#V  & \#F  & \#P  & R$_{s}$ & HD     & LFD  &  & \#V  & \#F & \#P & R$_{s}$ & HD     & LFD &                    \\ \midrule

PolyFit~\cite{nan2017polyfit}        
& 84   & 72  & \textbf{11}     &40\%   & 0.0473 & 4365   &
& 91   & 78   & \textbf{12}    & 15\%  & 0.0458 & 7779 &  
0.3\%    \\

KSR~\cite{bauchet2020kinetic}
& 280  & 97  & 97    &78\%   & 0.0397 & 5905   &
& 32   & 42  & 40     & 54\%   & 0.1131 & 8713 &  
1.1\%   \\

ProxyRecon~\cite{ProxyRecon24}
&107  & 114 & 58    &100\%   & 0.0243 & 4364   &  
& 60  & 90  & 34     & 100\%  & 0.0256 & 5340 &  
21.0\%    \\

\midrule

CA~\cite{yang2021unsupervised}
& \textbf{60}   & 180 & 180     & 100\%  & 0.0363 & 6246   &  
& 56   & 168  & 168     & 100\%  & 0.0396 & 6987  &
0.1\%   \\

BSP-Net~\cite{chen2020bsp}     
& 132    & 96   & 84    &100\%   & 0.0431      &6671    & 
& 102    & 170    & 67     & 100\%  & 0.0487      & 7162  &  
0.9\%  \\

\midrule
Ours                    
& 64   & \textbf{36}  & 14    &100\%   & \textbf{0.0154} &  \textbf{3873}   & 
& \textbf{27}   & \textbf{32}  & 15  &100\%  & \textbf{0.0219} & \textbf{4932} & 
\textbf{76.7\%}  \\ 
\bottomrule

\end{tabular}
}
\caption{
Quantitative comparison of our method with state-of-the-art (SOTA) methods and user study results. We report geometric properties (number of vertices \#V, faces \#F, and planes \#P) and distance metrics (Hausdorff distance = HD, Light Field Distance = LFD), which evaluate the structural simplicity, similarity to the reference, and the ratio of successful outputs (R$_s$), respectively.
}
\label{tab:comparison}
\end{table*}

\vspace{-10pt}

\paragraph{Training detail.}
The model is trained for 100k steps with a batch size of 128, using the AdamW optimizer (weight decay \( \lambda = 0.1 \), \( \beta_1 = 0.9 \), \( \beta_2 = 0.95 \)). The learning rate schedule includes a 5k-step warm-up phase, where the learning rate linearly increases from \(10^{-7}\) to \(10^{-4}\), followed by a cosine decay back to \(10^{-7}\). 
The point cloud encoder adopts a ResNet-style~\cite{he2016deep} architecture with sparse 3D convolutions~\cite{spconv2022}, extracting \( \mathbb{R}^{512} \) features in \(4^3\) from the input \(128^3\) voxel space. 
The decoder is a classical transformer~\cite{vaswani2017attention} with 12 layers, each featuring 8 attention heads, a model dimension \(d_{\text{model}} = 512\), and a feed-forward dimension \(d_{\text{ff}} = 2048\). 
All experiments were conducted on a server equipped with 8 NVIDIA RTX 4090 GPUs.

\begin{figure*}[tbp]
    \centering
    \vspace{-2mm}
    \includegraphics[width=\textwidth]{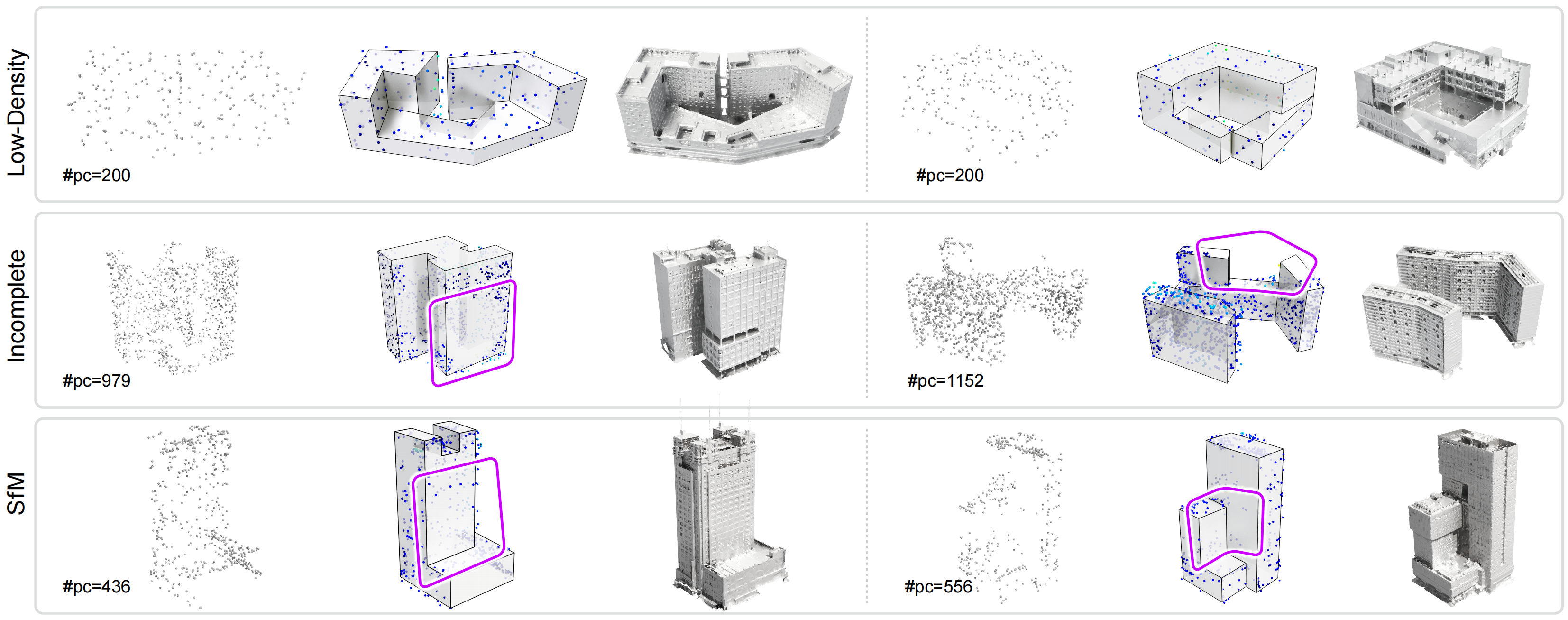}
    \caption{Vsualization examples on diverse low-quality input point clouds: low-density (\#pc=200) and incomplete point clouds sampled from the dense meshes, as well as SfM point clouds where both issues often coexist.}
    \label{fig:diverse_input}
    \vspace{-3mm}
\end{figure*}

\vspace{-10pt}

\paragraph{Data augmentation.}

To enhance the model's ability to handle low-quality point clouds from real-world scenarios, we perform sparsity sampling on the mesh during training data synthesis and post-process the sampled point clouds to introduce non-uniformity, incompleteness, and noise. The number of points is randomly chosen from the range \([200, 2000]\).
Specifically, given the target number of samples \(N\), we first generate a clean mesh and then apply non-uniform sampling to create \(5 \times N\) points, where non-uniformity is introduced through random weights on the triangular faces. Incompleteness is introduced by randomly selecting anchor points and iteratively dropping nearby points, until \(N\) points remain. Finally, we add uniform or Perlin noise~\cite{perlin1985image} to the spatial positions of these points. By pairing the low-quality sampled point clouds with programs as training data, we aim to improve the model's robustness in handling real-world data imperfections.

\subsection{Evaluation}

\paragraph{Evaluation datasets.}

We collected two test datasets. The first comprises 270 building instances from point clouds generated via structure-from-motion (SfM) using UAV imagery, with corresponding MVS dense meshes as reference. To comprehensively evaluate our method, we additionally collected 1,038 building instances in MVS dense mesh from the UrbanBIS~\cite{UrbanBIS}. Sparse point clouds were sampled from these meshes at an approximate density of one point per 20$m^2$, with potential noise and incompleteness originating from the quality of the original meshes.
Notably, the second dataset allows for decoupled analyses of various low-quality conditions by adjusting sampling methods.

\vspace{-10pt}

\paragraph{Metrics.}
We compare our method with five existing approaches for point cloud abstraction. To evaluate their ability to extract structured 3D representations from unstructured points, we use Hausdorff Distance (HD) and Light Field Distance (LFD) to measure geometric and visual errors between the abstraction and the reference. In both the SfM and sparse settings, the mesh from MVS dense reconstruction serves as the reference.
We also report the geometric properties of the abstraction, including the number of vertices (\#V), faces (\#F), and planes (\#P), as well as the ratio of successful outputs (R$_s$) on the evaluation dataset. Quantitative results in Table~\ref{tab:comparison} demonstrate that our method robustly handles various inputs,
achieving the best performance in nearly all geometric properties and distance metrics. Notably, some methods achieve better geometric properties at the cost of lower success rates, as their metrics are based only on simpler successful cases.

\vspace{-10pt}

\paragraph{Comparing to traditional methods.}
We compare our method with three optimization-based approaches for architecture proxy reconstruction: KSR~\cite{bauchet2020kinetic}, PolyFit~\cite{nan2017polyfit}, and ProxyRecon~\cite{ProxyRecon24}. PolyFit and KSR use planes as primitives to create proxy models, relying on plane detection algorithms like RANSAC~\cite{ransac} or Region Growing. These algorithms struggle with sparse, unstructured point clouds, resulting in suboptimal performance for KSR and PolyFit. In our experiments, we set the plane detection parameters to a maximum point-to-plane distance of 1.5m, a minimum of 10 supporting points, and an angle threshold of 40$^\circ$.
These methods often fail either due to plane detection failure (see Supp.) or computation times exceeding 40 minutes.
ProxyRecon~\cite{ProxyRecon24} uses convex hulls as primitives to create architectural proxies. This representation inherently limits the ability to model non-convex structures, such as U-shaped buildings or those with multiple branches.

\vspace{-10pt}

\begin{figure*}[tbp]
    \centering
    \includegraphics[width=0.97\textwidth]{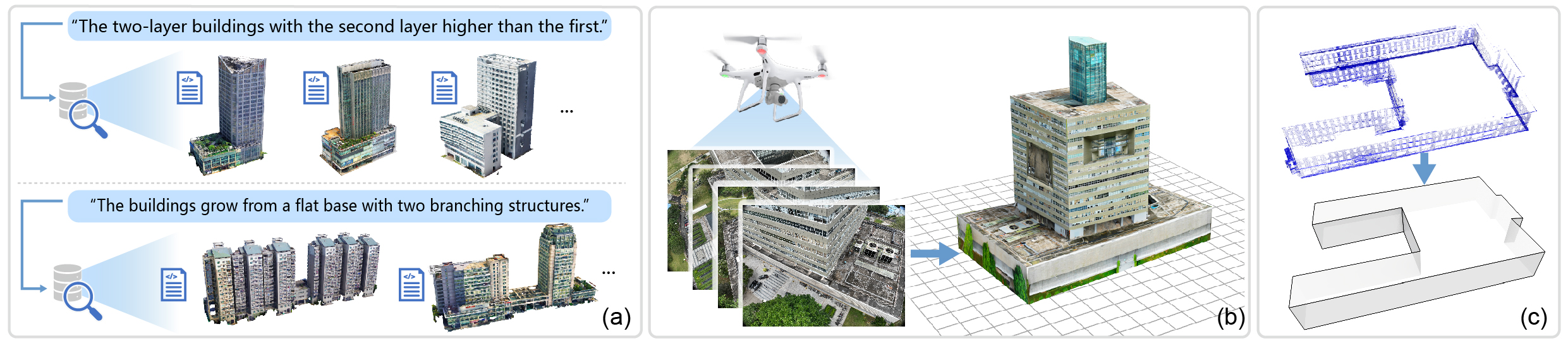}
    \caption{More applications of ArcPro: (a) architecture geometry structure analysis and natural language retrieval; (b) processing of raw SfM point clouds with ground points from drone aerial multi-view images; (c) processing of LiDAR-derived point clouds.}
    \vspace{-10pt}
    \label{fig:applications}
\end{figure*}

\paragraph{Comparing to learning-based methods.}

We compare ArcPro to two other learning-based methods for 3D structure abstraction from point clouds: CA~\cite{yang2021unsupervised}, which uses cubes as geometric primitives, and BSP-Net~\cite{chen2020bsp}, which relies on planes.
We train both methods on our synthetic dataset.
The quantitative and visual comparisons in Figure~\ref{fig:comparison} demonstrate that our structured representation outperforms theirs in handling architectural inputs.
CA, limited to fitting data with boxes, often struggles to accurately capture complex and diverse architectural structures. BSP-Net offers the most flexible structural representation among the three, but this flexibility leads to overfitting the noise in the input. Our proposed structured representation strikes a balance between the two, capturing the main structure from the input without overfitting the noise.


\vspace{-10pt}

\paragraph{Analysis.}
We control different sampling methods on the second type of evaluation dataset, that is, the MVS dense mesh, to obtain various low-quality sampled point clouds for analysis.
We investigate two relatively decoupled and common types of low-quality data: low-density and incompleteness, which often coexist in SfM point clouds. The results are shown in Figure~\ref{fig:diverse_input}.
For low-density point clouds, even with as few as 200 points, our method is still able to infer the underlying structure that aligns with the reference mesh. This capability stems from our DSL design, which constrains the solution space, along with data biases introduced during training that guide the network to favor predictions with clean, balanced geometry.
For cases with severe incompleteness, some lead to structural ambiguity, while others do not. In the left example, despite a large incomplete area in the plane, intact surrounding regions provide clues that help preserve the underlying structure. In contrast, the right example's incompleteness creates ambiguity; in this case, our method tends to adhere closely to the input point cloud rather than attempting to complete it.

\vspace{-10pt}

\paragraph{User study.}
We conducted a user study with 110 participants to evaluate the quality of abstractions generated by various methods. In this study, 20 {\em randomly\/} selected models were used, and the user study participants selected the best abstraction from six results--each generated by a different method--based on achieving a balance between fidelity and simplicity. Detailed results are shown in the last column of Table~\ref{tab:comparison}. Our method received a preference rating of 76.7\%, making it the most preferred approach. This confirms the effectiveness of our method for abstracting point clouds, as our results are not only geometrically and visually accurate but also aesthetically appealing.

\section{Conclusion, Limitation, and Future Work}
\label{sec:future}

We present ArcPro, a program-based learning framework to recover structured 3D abstractions from low-quality, unstructured building point clouds.
By designing a DSL, we transform the solution space into a compact program space that embeds architectural structure priors, capturing a wide range of building abstractions.
Extensive experiments show that our method outperforms other SOTA methods, and effectively handles diverse low-quality point clouds.

\vspace{-10pt}

\paragraph{Applications.}

We explore the application potentials of our method, as showcased in Figure~\ref{fig:applications}.
ArcPro bridges program-level information processing with 3D building model representation. Its natural language features facilitate architectural analysis and retrieval, while its editability, interpretability, and scalability support diverse statement types.
Another promising area is the processing of multi-view aerial images without relying on point cloud segmentation by incorporating ground point simulation into data augmentation. Compared to traditional MVS, our approach provides a significant advantage in inference speed, while generating lightweight, textured 3D abstractions.

\vspace{-15pt}

\paragraph{Limitations.}
ArcPro lacks precision in capturing detailed structures, likely due to its reliance solely on next-token prediction loss. While this approach aids early low-frequency structure learning, it struggles with high-frequency detail.
Also, structure recovery from sparse point clouds may have multiple valid solutions, as shown by the second incomplete point cloud in Figure~\ref{fig:diverse_input}. Our method currently infers only a single solution via top-1 sampling, and using top-3 sampling reduces output quality rather than improving diversity.

\vspace{-15pt}

\paragraph{Future work.}
We shall focus on scalability, diverse data modalities, and targeted designs. With program scalability, we can define new statements for more geometric features, such as curved surfaces or sloped roofs. Our framework is not limited to point clouds as input; it may be extended to images or multi-modal program learning. Currently, we use a standard network with only next-token prediction, treating tokens generally from a network perspective. This leaves room for targeted designs, such as embedding explicit geometric information to enhance 3D modeling.

\vspace{-15pt}
\paragraph{Acknowledgments.}
This work was supported in parts by National Key R\&D Program of China (2024YFB3908500), NSFC (U21B2023), ICFCRT (W2441020), GD Basic and Applied Basic Research Foundation (2023B1515120026), DEGP  (2022KCXTD025), Shenzhen Science and Technology Program (KJZD20240903100022028, KQTD202108 11090044003, RCJC20200714114435012),
and Scientific Development Funds from Shenzhen University.

{
    \small
    \bibliographystyle{ieeenat_fullname}
    \bibliography{main}
}

\appendix
\clearpage
\setcounter{page}{1}
\maketitlesupplementary


\section{Experiments}

\paragraph{Network structures.} 
The point cloud encoder is a ResNet-style architecture built with sparse 3D convolutions. It processes voxelized point clouds of size \(128^3\) and progressively reduces the spatial resolution through five stages, each consisting of residual blocks with sparse convolutions. The network begins with a basic sparse convolutional block and follows a structure where feature dimensions increase across stages: \( [64, 128, 192, 384, 512] \). Each stage employs two residual blocks, with downsampling implemented using sparse max pooling. The encoded output is compressed into a \( \mathbb{R}^{512} \) feature representation, corresponding to a spatial resolution of \(4^3\). 
The program decoder is a classical transformer with 12 layers, each featuring 8 attention heads, a model dimension \(d_{\text{model}} = 512\), and a feed-forward dimension \(d_{\text{ff}} = 2048\).
The encoded point cloud features are incorporated into the program decoder via cross-attention, enabling effective conditional program generation that aligns with the input point cloud.

\paragraph{Training details.} The input point cloud is normalized to fit within a unit cube \([-0.5, 0.5]\) by centering and scaling its coordinates based on the data's range. For tokenization, numerical values such as coordinates or height values are discretized within the range \([-1, +1]\). This range is divided into intervals with a resolution of 0.02, resulting in 100 distinct numeric tokens to represent the corresponding discrete values.
We use a label smoothing strategy: non-numeric tokens have a ground truth probability of 0.95, with 0.05 distributed evenly among others; numeric tokens have 0.5, with 0.25 assigned to adjacent tokens to reflect their continuous nature for better optimization.

\begin{figure*}[tbp]
    \centering
    \includegraphics[width=0.9\textwidth]{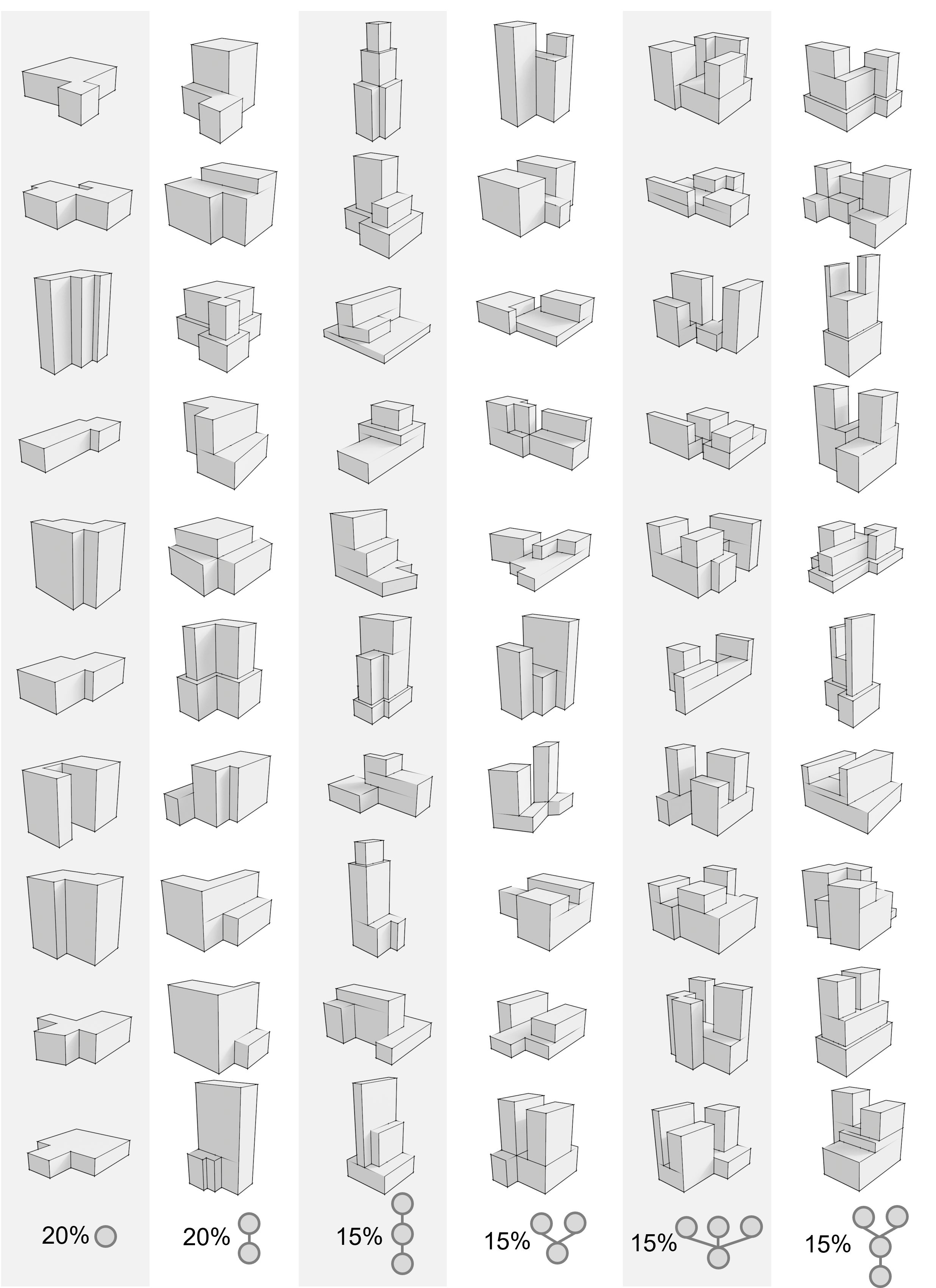}
    \caption{Synthetic training data by procedural generation. The bottom row shows six architecture tree modes and their sampling ratios.}
    \label{fig:synthetic}
\end{figure*}

\begin{figure*}[tbp]
    \centering
    \includegraphics[width=\textwidth]{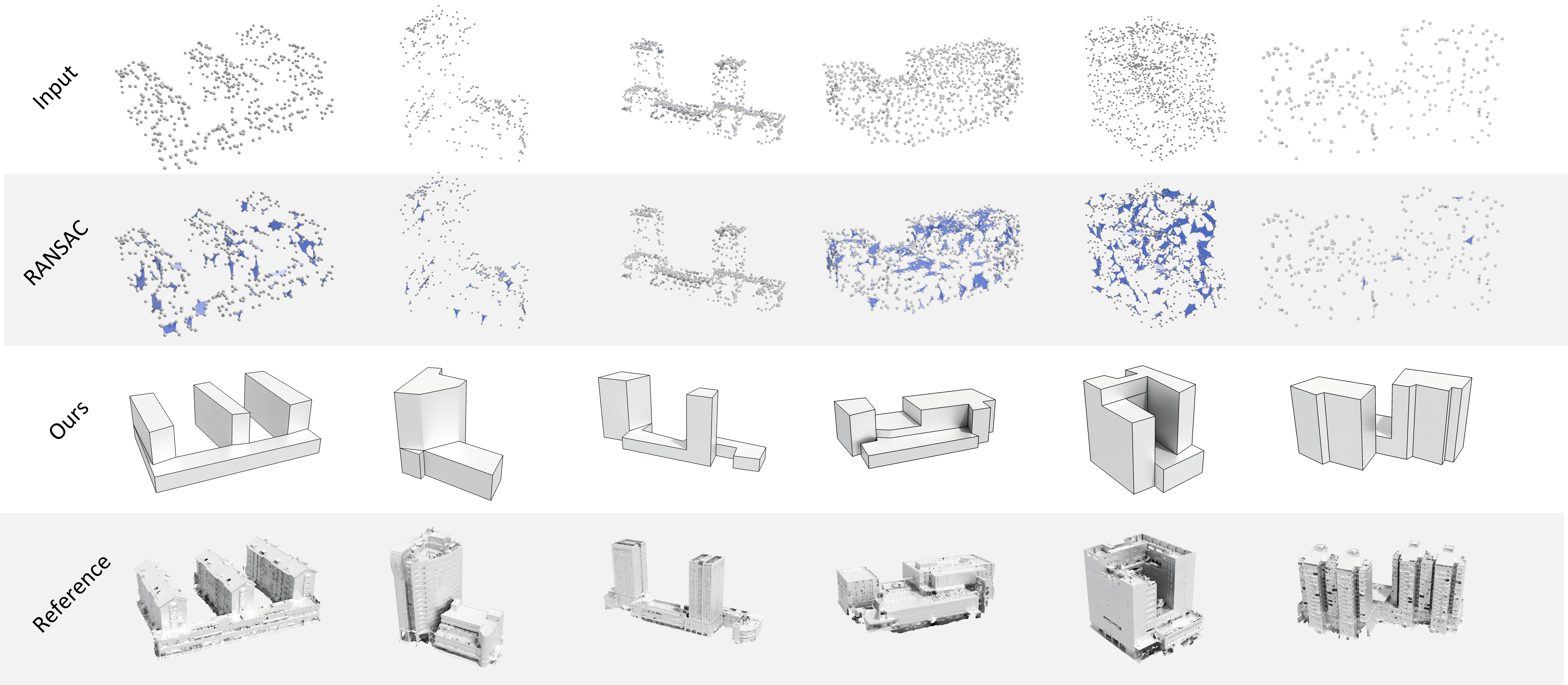}
    \caption{
    RANSAC plane detection results on the input from Figure 5 in the main paper. The results demonstrate that RANSAC struggles with diverse and low-quality architecture point clouds, leading to the failure of traditional methods that rely on RANSAC.
    }
    \label{fig:ransac}
\end{figure*}

\begin{figure*}[tbp]
    \centering
    \includegraphics[width=0.95\textwidth]{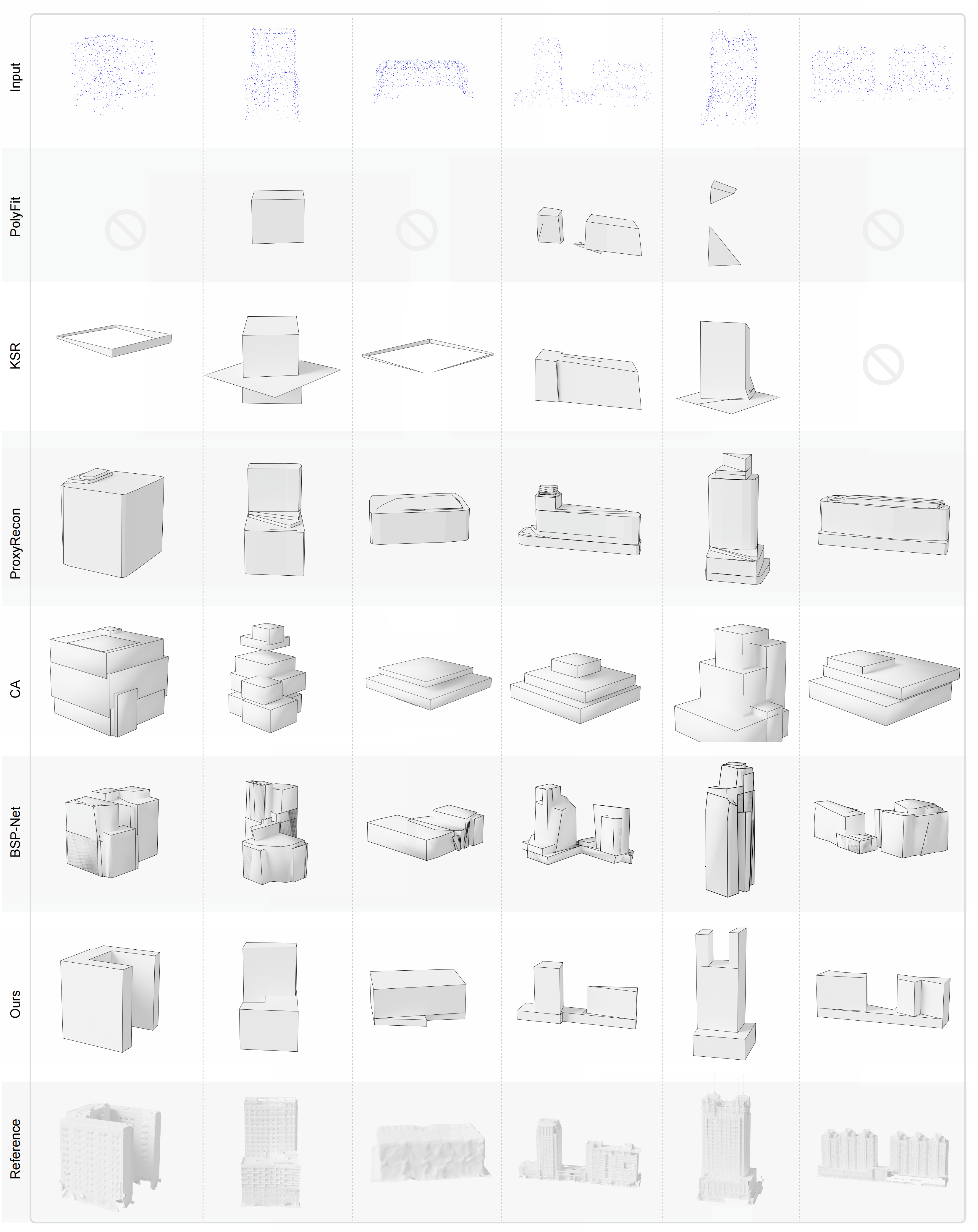}
    \caption{The user study examples comparing our method to other methods.}
    \label{fig:us0}
\end{figure*}

\begin{figure*}[tbp]
    \centering
    \includegraphics[width=0.95\textwidth]{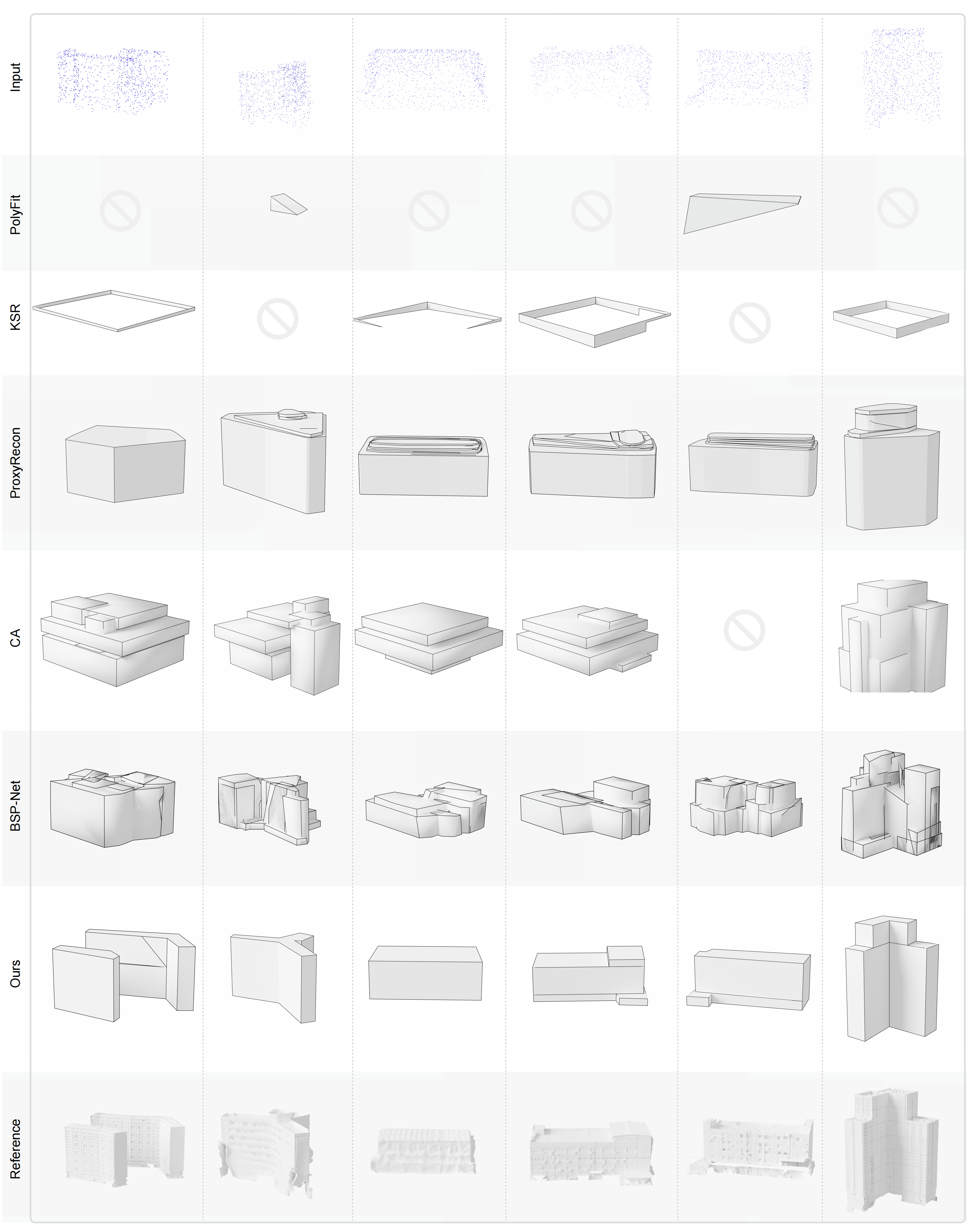}
    \caption{The user study examples comparing our method to other methods.}
    \label{fig:us1}
\end{figure*}

\begin{figure*}[tbp]
    \centering
    \includegraphics[width=0.95\textwidth]{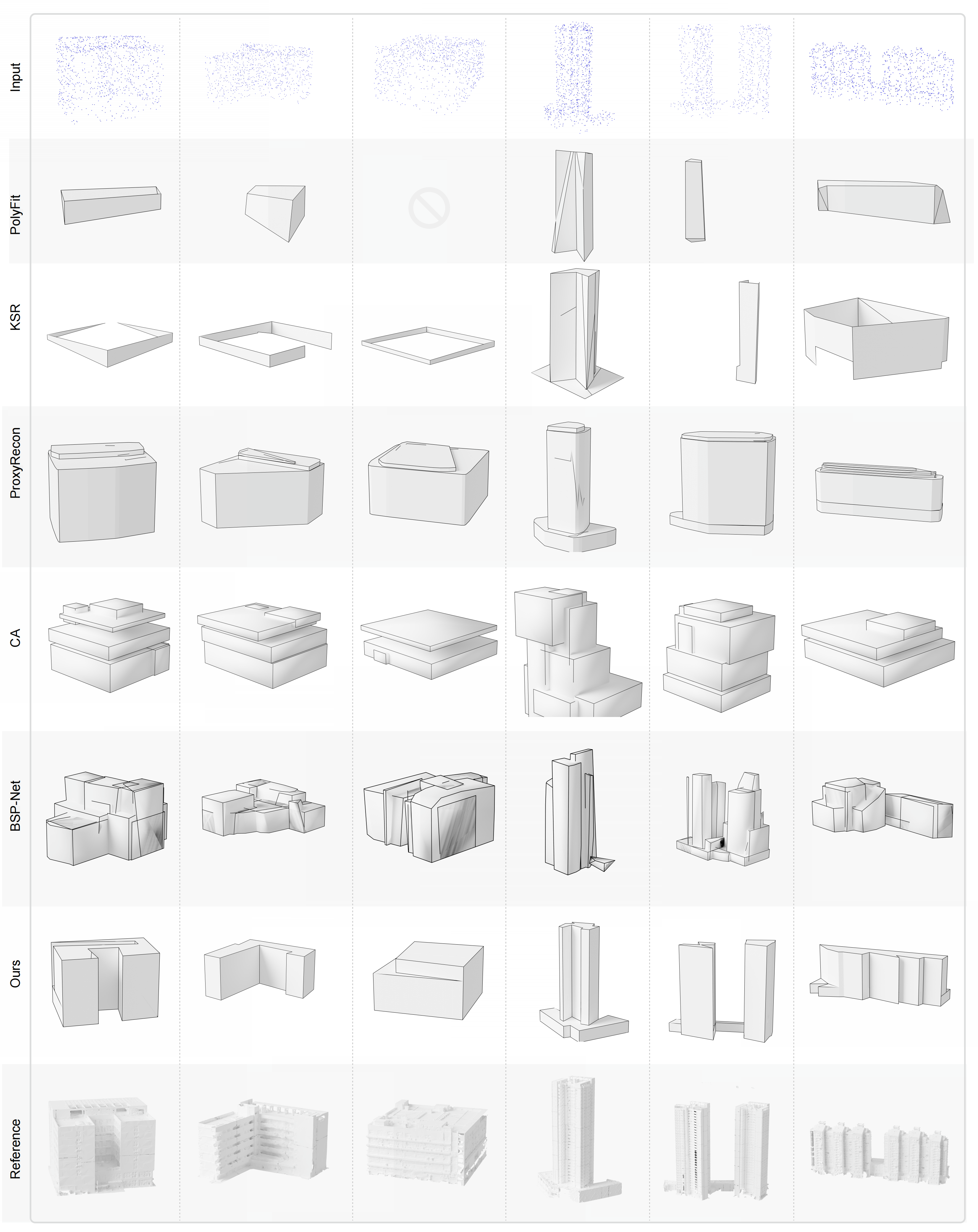}
    \caption{The user study examples comparing our method to other methods.}
    \label{fig:us2}
\end{figure*}

\paragraph{Additional illustrations.}
We present examples of procedurally generated synthetic training data, as illustrated in Figure~\ref{fig:synthetic}. These are generated online during training, including six types of architectural tree models, which are sampled based on specific proportions.
More results of our method applied to Structure-from-Motion (SfM) point clouds and sparse sampling point clouds are provided, as shown in Figure~\ref{fig:comparison}.
Furthermore, we examine the performance of RANSAC plane detection on low-quality point clouds derived from the experimental section of the main paper.
As shown in Figure~\ref{fig:ransac}, these results reveal RANSAC's struggles with sparse point clouds, causing traditional methods to fail.
Finally, we present user study examples comparing our method to alternative approaches. These examples are shown in Figures~\ref{fig:us0}, \ref{fig:us1}, and \ref{fig:us2}.

\begin{table}[t]
    \caption{
        Ablation study for training configuration.
    }
    \vspace{-10pt}
    \label{tab:ablation_study}
    \centering
    \setlength{\tabcolsep}{5pt}
    \resizebox{\linewidth}{!}{
    \begin{tabular}{@{}ccc|cc|cc@{}}
        \toprule
        \multirow{2}{*}{\makecell{Noise\\Scale}} &
        \multirow{2}{*}{\makecell{Incomplete\\Ratio}} &
        \multirow{2}{*}{\makecell{$\langle\text{/p}\rangle, \langle\text{/h}\rangle$\\Token}} & 
        \multicolumn{2}{c|}{SfM} & \multicolumn{2}{c}{Sparse Sampling} \\
        \cmidrule(lr){4-5} \cmidrule(lr){6-7}
        & & & HD (\textdownarrow) & LFD (\textdownarrow) & HD (\textdownarrow) & LFD (\textdownarrow) \\ 
        \midrule

        0 & \multirow{2}{*}{} & \multirow{2}{*}{} & 0.0195 & 4192 & 0.0291 & 5387 \\
        0.05 &                   &                    & 0.0177 & 4338 & 0.0237 & 4958 \\
        \midrule
        \multirow{2}{*}{} & 0\% & \multirow{2}{*}{} & 0.0169 & 4210 & 0.0250 & 5033 \\
                           & $[50\%, 90\%]$ &                  & 0.0187 & 4396 & 0.0266 & 5211 \\
        \midrule
         &  & w/o & 0.0181 & 4259 & 0.0272 & 5212 \\
        \midrule
        0.02 & $[10\%, 50\%]$ & w/ & \textbf{0.0154} & \textbf{3873} & \textbf{0.0219} & \textbf{4932} \\
        \bottomrule
    \end{tabular}
    }
\end{table}

\paragraph{Ablation study.}
See Table~\ref{tab:ablation_study}.
For data augmentation, both \textit{noise scale} and \textit{incomplete ratio} should be moderate: if too weak, they fail to adequately simulate the low-quality nature of real point clouds;  if too strong, the problem becomes overly ill-posed, degrading performance and destabilizing training. 
For \textit{token schema}, we use \(\langle\text{/p}\rangle\) and \(\langle\text{/h}\rangle\) as end tokens for point coordinates and height values, respectively.
While parsing works without them, they improve performance and stabilize training.
For \textit{geometry refinement}, omitting it during inference has little impact on the metrics but noticeably degrades visualization due to slight misalignment of points and lines.

\begin{figure}[h]
\centering
\begin{minipage}[h]{0.5\linewidth}
    \centering
    \caption{Generalization.}
    \vspace{-10pt}
    \includegraphics[width=4cm]{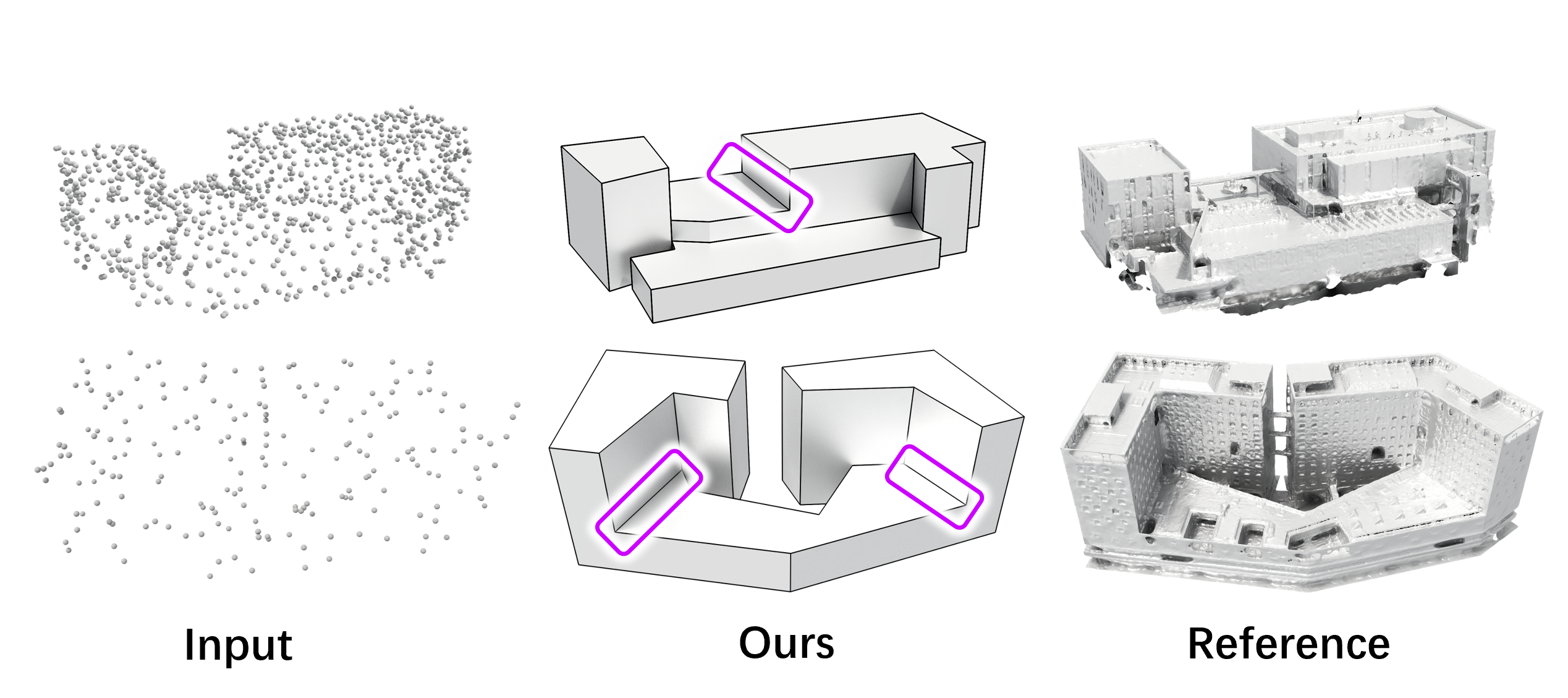}
    \label{fig:generalization}
\end{minipage}%
\hfill
\begin{minipage}[h]{0.5\linewidth}
    \centering
    \setlength{\tabcolsep}{5pt}
    \captionof{table}{Robust to data ratio.}
    \vspace{-10pt}
    \resizebox{\linewidth}{!}{
        \begin{tabular}{@{}c|cc|cc@{}}
            \toprule
            \multirow{2}{*}{\makecell{Training data\\(4-gon : 6-gon)}} &
            \multicolumn{2}{c|}{4-gon} & \multicolumn{2}{c}{6-gon} \\
            \cmidrule(lr){2-3} \cmidrule(lr){4-5}
            & \#n & HD  & \#n & HD \\ 
            \midrule
            20\% : 80\% & 4.07 & 0.0089 & 5.95 & 0.0085 \\
            50\% : 50\% & 4.03 & 0.0091 & 5.94 & 0.0089 \\
            80\% : 20\% & 4.04 & 0.0087 & 5.95 & 0.0090 \\
            \bottomrule
        \end{tabular}
    }
    \label{tab:robustness}
\end{minipage}
\end{figure}

\paragraph{Generalizability and robustness.}
Our goal is to learn \textit{conditional} mapping from input point clouds, where domain shifts between synthetic and real data can be 
mitigated since the input provides a contextual hint during inference.
We cannot retrieve the most similar shape from the training set due to online data synthesis, but Figure~\ref{fig:generalization} shows that \textit{our method can infer unsynthesized or unseen shapes}.
According to our procedural rules (see Sec 4.2), when $M > 1$, each edge should lie on the extension of a parent edge, but this is not satisfied in Figure~\ref{fig:generalization} above.
We also explored robustness against varying data ratios by preparing two test sets (4-gon and 6-gon) and three training sets with different mixing ratios; see Table~\ref{tab:robustness}.

\newpage

\section{Applications}
\begin{figure*}[tbp]
    \centering
    \includegraphics[width=\textwidth]{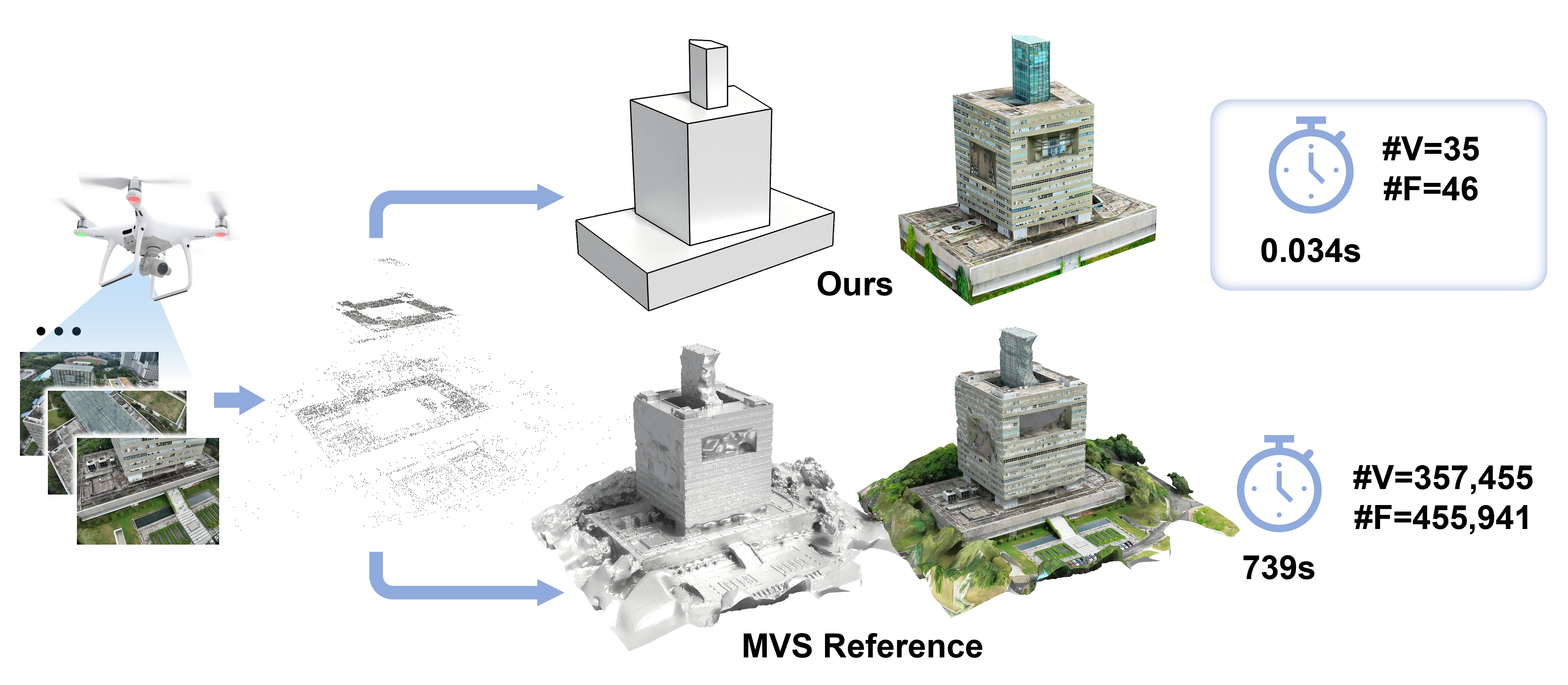}
    \caption{The result processes raw SfM point clouds from multi-view aerial images, bypassing building segmentation. Compared to traditional MVS methods, ArcPro significantly enhances inference speed while generating lightweight, textured 3D abstractions.}
    \label{fig:mvs}
\end{figure*}

\paragraph{Multi-view aerial images.}
We extend our framework to process raw SfM point clouds from multi-view aerial images, bypassing building segmentation. This introduces noise like ground points and outliers. To mitigate this, we augment data by expanding a building's footprint's convex hull or bounding box to simulate ground and adding noise to represent trees, cars, and other elements.
This allows us to more effectively process unsegmented SfM point clouds. Compared to traditional MVS methods, ArcPro significantly improves inference speed while producing lightweight, textured 3D abstractions, as shown in Figure~\ref{fig:mvs}.
ArcPro takes 0.034s on an RTX 4090 GPU, compared to 739s for the traditional MVS pipeline (using the commercial software ContextCapture), achieving a 10,000x reduction in time and data size (\#V for vertices, \#F for triangular faces). This allows faster processing, lower rendering costs, and more efficient data transfer and storage, which is critical for spatial computing applications.

\begin{figure*}[tbp]
    \centering
    \includegraphics[width=\textwidth]{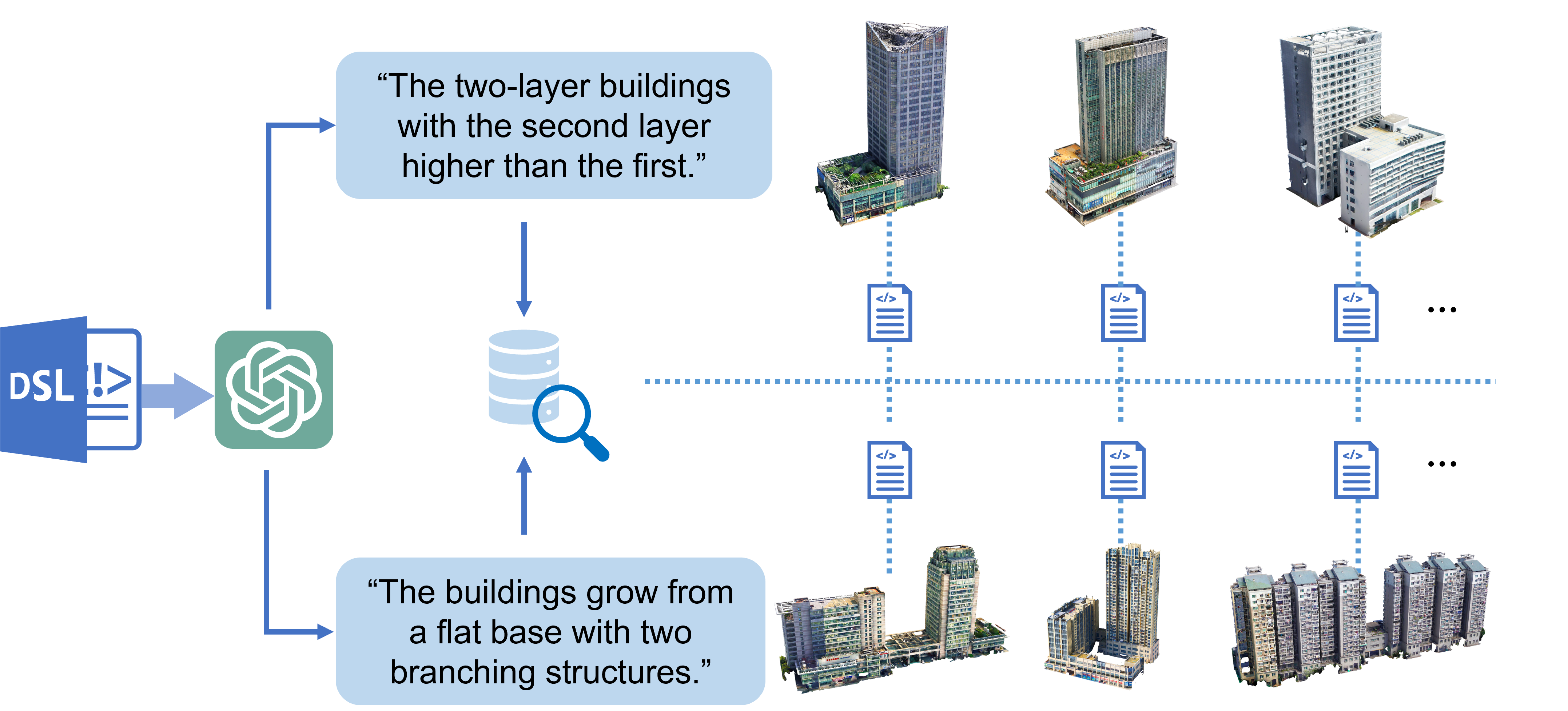}
    \caption{Architecture geometry structure analysis and natural language retrieval. Prompting ChatGPT with DSL definitions to convert geometric structure queries into Python code \texttt{IsMatching(program)} to vertify each program for retrieving matching programs.}
    \label{fig:nlp}
\end{figure*}

\vspace{-10pt}
\paragraph{Natural language retrieval.}
Our method encodes architectural structures as programs, leveraging their linguistic properties for natural language-driven analysis and retrieval using large language models prompt by DSL definition.
ArcPro transforms a database of 3D architectural models into corresponding programmatic representations, establishing connections between programs and 3D models.
For example, as shown in Figure~\ref{fig:nlp}, given a query like \textit{“two-layer buildings where the second layer is higher than the first,”} a language model such as ChatGPT will generate Python code for an \texttt{IsMatching(program)} function based on the DSL definition, implementing the logic to verify each program. The function returns \texttt{True} for programs meeting the criteria and \texttt{False} otherwise, enabling the retrieval of relevant 3D architectural models effectively.

\begin{figure*}[tbp]
    \centering
    \includegraphics[width=\textwidth]{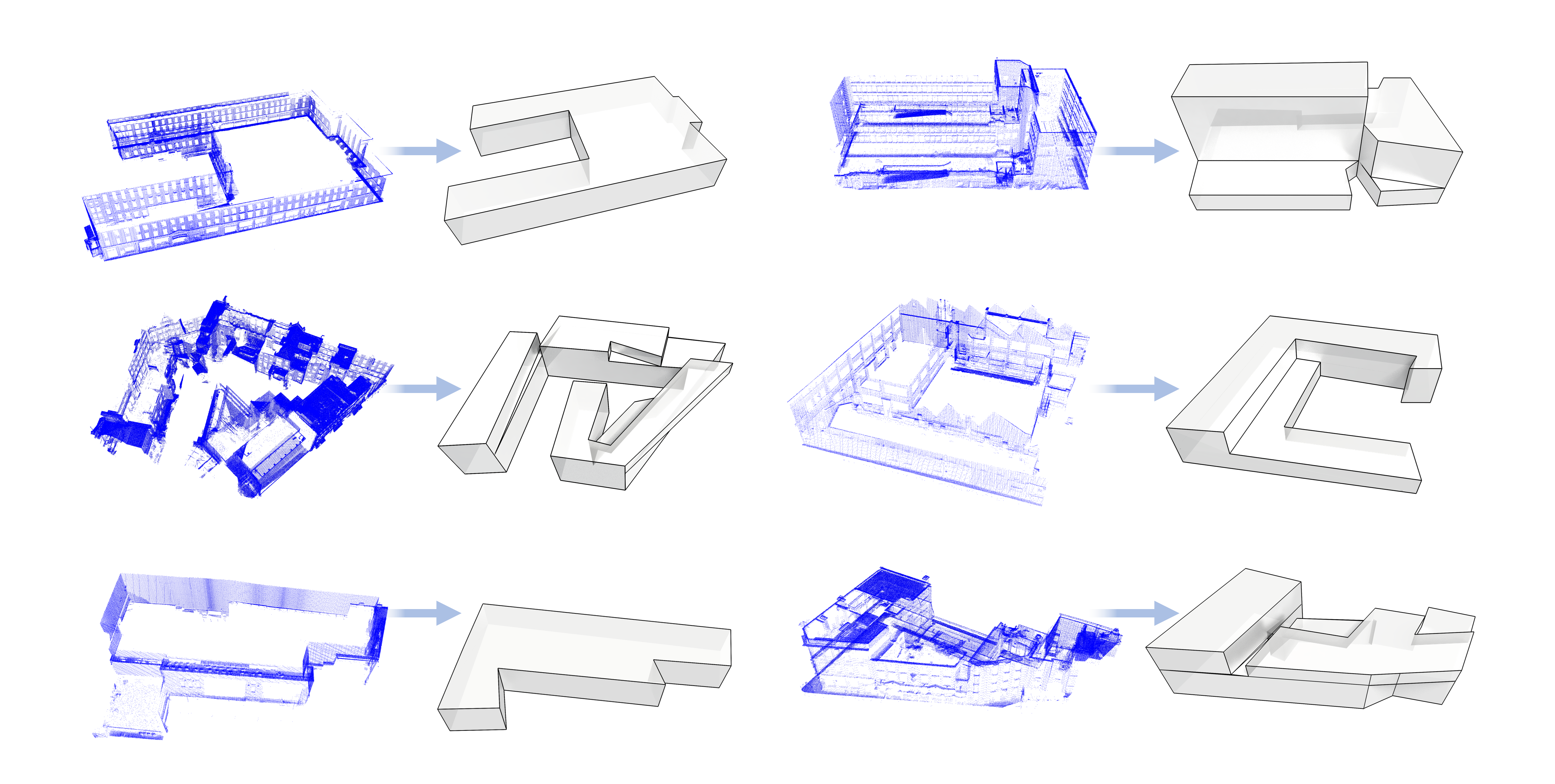}
    \caption{Results with LiDAR point clouds. Without specialized data augmentation, our method achieved reasonable performance.}
    \label{fig:lidar}
\end{figure*}

\vspace{-10pt}
\paragraph{LiDAR point clouds.}
We also explore the performance of ArcPro on LiDAR point cloud input, using data from the DublinCity dataset~\cite{zolanvari2019dublincity}, as shown in Figure~\ref{fig:lidar}. Even without incorporating specific design in data augmentation to simulate the characteristics of LiDAR point clouds, our method is still able to achieve reasonable performance.

\begin{figure}[t]
    \centering    \includegraphics[width=0.48\textwidth]{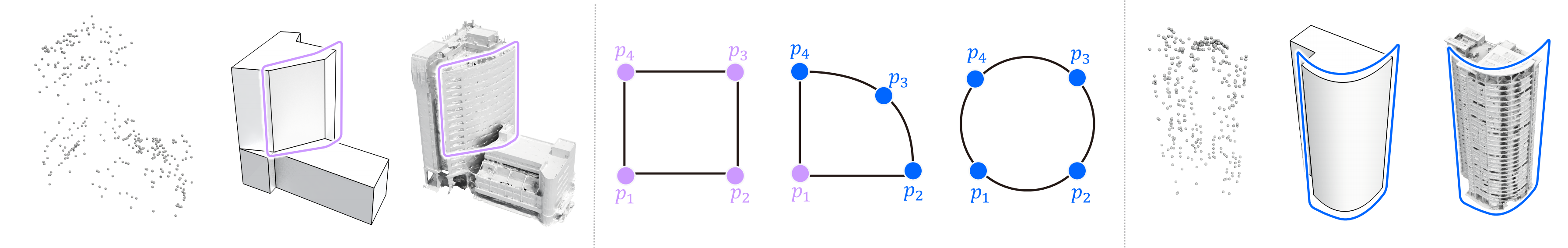}
    \caption{The examples of non-planar surfaces in the current and extended ArcPro framework.}
    \label{fig:non_planer}
\end{figure}

\vspace{-10pt}
\paragraph{Non-planar surfaces.}
As our work primarily focuses on recovering planar surfaces, curved surfaces, such as the one shown in Fig.~\ref{fig:non_planer} left, need to be approximated by polygonal contours. Extending our framework to handle non-planar contours is quite straightforward. By distinguishing curve points from corner points at the token level (marked in purple or blue), the geometry compiler can fit curve segments as parametric curves.
We synthesize corresponding training data to obtain preliminary results shown in Fig.~\ref{fig:non_planer} right, highlighting the potential of our program framework.

\section{Motivation of DSL}
We design our DSL to align with architectural priors and be syntactically compatible with traditional procedural building generation (PBG), instead of using a more general shape language.
This approach offers these advantages:
\begin{itemize}
    \item \textit{More compact representation} with a more efficient solution space. For example, unlike sketch-and-extrude, which requires six DoF (origin and orientation), our approach employs a parent layer index to simultaneously specify the 3D coordinate frame and layer hierarchy.
    \item \textit{Leveraging mature PBG research} for large-scale training data synthesis, where architectural priors can be injected.
    \item \textit{Extensibility} to accommodate new statements that support additional architectural features, such as roof structures from OpenStreetMap (OSM) or for-loops for repetitive elements like windows in façade modeling.
    \item \textit{Explicit encoding of building properties}, such as hierarchical relationships in \texttt{CreateLayer} statements, facilitating language-based retrieval and analysis.
    \item \textit{Editability} through parametric modeling and the relation of geometric elements align with architectural features.
\end{itemize}

\begin{figure*}[tbp]
    \centering
    \includegraphics[width=0.95\textwidth]{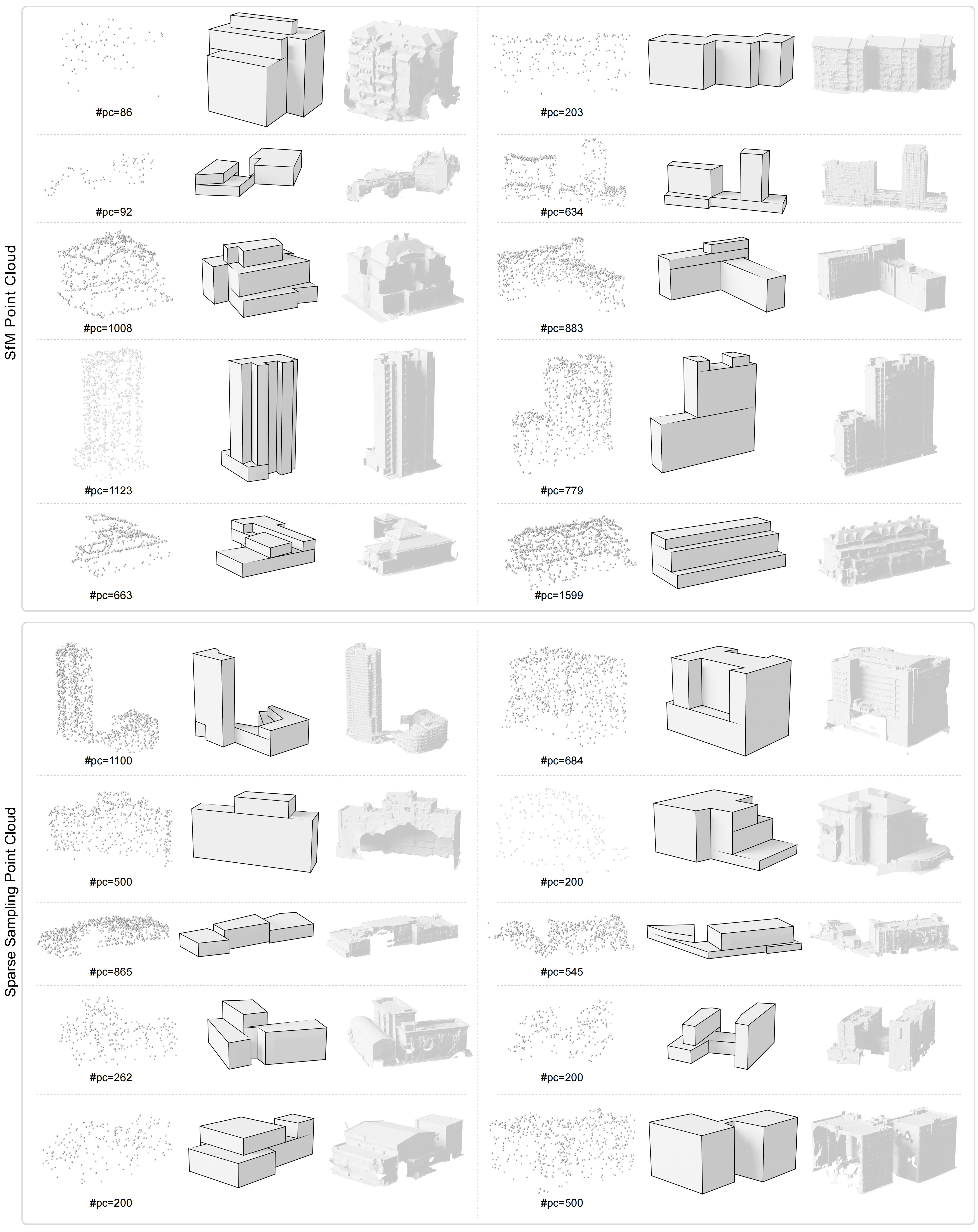}
    \caption{More results of our method applied to Structure-from-Motion (SfM) point clouds and sparse sampling point clouds. Our method can recover structured 3D abstractions from low-quality architectural point clouds that are non-uniform, incomplete, and noisy.}
    \label{fig:comparison}
\end{figure*}

\end{document}